\documentclass[]{aa}
\usepackage{txfonts}
\usepackage{graphicx}
\begin{document}
\title{The Cosmic Microwave Background Anisotropy Power Spectrum
  measured by Archeops}

   \subtitle{}

\author{ 
A.~Beno{\^\i}t\inst{1} \and 
P.~Ade \inst{2} \and 
A.~Amblard \inst{3, \, 24} \and 
R.~Ansari \inst{4} \and 
{{\'E}}.~Aubourg \inst{5, \, 24} \and
S.~Bargot \inst{4} \and
J.~G.~Bartlett \inst{3, \, 24} \and 
J.--Ph.~Bernard \inst{7,\, 16} \and 
R.~S.~Bhatia \inst{8} \and 
A.~Blanchard\inst{6} \and 
J.~J.~Bock \inst{8, \, 9} \and
A.~Boscaleri \inst{10} \and 
F.~R.~Bouchet \inst{11} \and 
A.~Bourrachot \inst{4} \and 
P.~Camus \inst{1} \and 
F.~Couchot \inst{4} \and
P.~de~Bernardis \inst{12} \and 
J.~Delabrouille \inst{3, \, 24} \and
F.--X.~D{\'e}sert \inst{13} \and 
O.~Dor{{\'e} \inst{11}} \and 
M.~Douspis \inst{6, \, 14} \and 
L.~Dumoulin \inst{15} \and 
X.~Dupac \inst{16} \and
P.~Filliatre \inst{17} \and 
P.~Fosalba \inst{11} \and
K.~Ganga \inst{18} \and 
F.~Gannaway \inst{2} \and 
B.~Gautier \inst{1} \and 
M.~Giard \inst{16} \and
Y.~Giraud--H{\'e}raud \inst{3, \, 24} \and 
R.~Gispert \inst{7\dag}\thanks{Richard Gispert passed away few weeks
after his return from the early mission to Trapani} \and 
L.~Guglielmi \inst{3, \, 24} \and
J.--Ch.~Hamilton \inst{3, \, 17} \and 
S.~Hanany \inst{19} \and
S.~Henrot--Versill{\'e} \inst{4} \and 
J.~Kaplan \inst{3, \, 24} \and
G.~Lagache \inst{7} \and 
J.--M.~Lamarre \inst{7,25} \and 
A.~E.~Lange \inst{8} \and 
J.~F.~Mac{\'\i}as--P{\'e}rez \inst{17} \and 
K.~Madet \inst{1} \and 
B.~Maffei \inst{2} \and
Ch.~Magneville \inst{5, \, 24} \and
D.~P.~Marrone \inst{19} \and
S.~Masi \inst{12} \and 
F.~Mayet \inst{5} \and 
A.~Murphy \inst{20} \and
F.~Naraghi \inst{17} \and 
F.~Nati \inst{12} \and
G.~Patanchon \inst{3, \, 24} \and
G.~Perrin \inst{17} \and 
M.~Piat \inst{7} \and 
N.~Ponthieu \inst{17} \and
S.~Prunet \inst{11} \and
J.--L.~Puget \inst{7} \and
C.~Renault \inst{17} \and 
C.~Rosset \inst{3, \, 24} \and
D.~Santos \inst{17} \and
A.~Starobinsky \inst{21} \and
I.~Strukov \inst{22} \and
R.~V.~Sudiwala \inst{2} \and 
R.~Teyssier \inst{11, \, 23} \and
M.~Tristram \inst{17} \and
C.~Tucker\inst{2} \and
J.--C.~Vanel \inst{3, \, 24} \and 
D.~Vibert \inst{11} \and 
E.~Wakui \inst{2} \and 
D.~Yvon \inst{5, \, 24}
}

   \offprints{reprints@archeops.org}
   \mail{benoit@archeops.org}
\institute{
Centre de Recherche sur les Tr{\`e}s Basses Temp{\'e}ratures,
BP166, 38042 Grenoble Cedex 9, France
\and
Cardiff University, Physics Department, PO Box 913, 5, The Parade,   
Cardiff, CF24 3YB, UK\and
Physique Corpusculaire et Cosmologie, Coll{\`e}ge de
France,  11 pl. M. Berthelot, F-75231 Paris Cedex 5, France
\and
Laboratoire de l'Acc{\'e}l{\'e}rateur Lin{\'e}aire, BP~34, Campus
Orsay, 91898 Orsay Cedex, France
\and
CEA-CE Saclay, DAPNIA, Service de Physique des Particules, 
Bat 141, F-91191 Gif sur Yvette Cedex, France
\and
Laboratoire d'Astrophysique de l'Obs. Midi-Pyr{\'e}n{\'e}es,
14 Avenue E. Belin, 31400 Toulouse, France
\and
Institut d'Astrophysique Spatiale, B{\^a}t.  121, Universit{\'e} Paris
XI,
91405 Orsay Cedex, France
\and
California Institute of Technology, 105-24 Caltech, 1201 East 
California Blvd, Pasadena CA 91125, USA
\and
Jet Propulsion Laboratory, 4800 Oak Grove Drive, Pasadena, 
California 91109, USA
\and
IROE--CNR, Via Panciatichi, 64, 50127 Firenze, Italy
\and
Institut d'Astrophysique de Paris, 98bis, Boulevard Arago, 75014 Paris,
France
\and
Gruppo di Cosmologia Sperimentale, Dipart. di Fisica, Univ.  ``La
Sapienza'', P. A. Moro, 2, 00185 Roma, Italy
\and
Laboratoire d'Astrophysique, Obs. de Grenoble, BP 53, 
 38041 Grenoble Cedex 9, France
\and
Nuclear and Astrophysics Laboratory, Keble Road, Oxford,  OX1 3RH, UK
\and
CSNSM--IN2P3, B{\^a}t 108, 91405 Orsay Campus, France
\and
Centre d'{\'E}tude Spatiale des Rayonnements,
BP 4346, 31028 Toulouse Cedex 4, France
\and
Institut des Sciences Nucl{\'e}aires, 53 Avenue des Martyrs, 38026
Grenoble Cedex, France
\and
Infrared Processing and Analysis Center, Caltech, 770 South Wilson
Avenue, Pasadena, CA 91125, USA
\and
School of Physics and Astronomy, 116 Church St. S.E., University of
Minnesota, Minneapolis MN 55455, USA
\and
Experimental Physics, National University of Ireland, Maynooth, Ireland
\and
Landau Institute for Theoretical Physics, 119334 Moscow, Russia
\and
Space Research Institute, Profsoyuznaya St. 84/32, Moscow, Russia 
\and
CEA-CE Saclay, DAPNIA, Service d'Astrophysique, Bat 709, 
F-91191 Gif sur Yvette Cedex, France
\and
F{\'e}d{\'e}ration de Recherche APC, Universit{\'e} Paris 7, Paris, France
\and
LERMA, Observatoire de Paris, 61 Av. de l'Observatoire, 75014 Paris, France
}

\date{Received 16 October 2002/Accepted 15 December 2002} 
   
   \abstract{ We present a determination by the Archeops experiment of
     the angular power spectrum of the cosmic microwave background
     anisotropy in 16 bins over the multipole range $\ell=15-350$.
     Archeops was conceived as a precursor of the Planck HFI
     instrument by using the same optical design and the same
     technology for the detectors and their cooling.  Archeops is a
     balloon--borne instrument consisting of a 1.5~m aperture diameter
     telescope and an array of 21 photometers maintained at $\sim
     100$~mK that are operating in 4 frequency bands centered at 143,
     217, 353 and 545~GHz. The data were taken during the Arctic night
     of February 7, 2002 after the instrument was launched by CNES
     from Esrange base (Sweden). The entire data cover $\sim 30$\% of
     the sky.  This first analysis was obtained with a small subset of
     the dataset using the most sensitive photometer in each CMB band
     (143 and 217~GHz) and 12.6\% of the sky at galactic latitudes
     above 30 degrees where the foreground contamination is measured
     to be negligible.  The large sky coverage and medium resolution
     (better than $15$~arcmin.) provide for the first time a high
     signal-to-noise ratio determination of the power spectrum over
     angular scales that include both the first acoustic peak and
     scales probed by COBE/DMR. With a binning of $\Delta \ell$=7 to
     25 the error bars are dominated by sample variance for $\ell$
     below 200. A companion paper details the cosmological
     implications.

         }

   \maketitle


\section{Introduction}

Observations of the Cosmic Microwave Background (CMB) temperature
anisotropies have provided answers to fundamental questions in
cosmology.  The observational determination of the CMB angular power
spectrum has already led to important insights on the structure and
evolution of the universe. Most notable are the conclusions that the
geometry of space is essentially flat~(\cite{toco,boom1,maxima1}) and
that the measurements are consistent with the inflationary
paradigm~(\cite{boom2,maxima2,dasi,cbi,vsa}). Since the first
detection of CMB anisotropy with COBE/DMR~(\cite{smoot}), a host of
experiments have measured the spectrum down to sub--degree scales, but
measurements at large angular scales remain difficult, due to the
large sky coverage required to access these modes. This difficulty
will be overcome by the future full--sky space missions MAP and
Planck.

This paper presents the first results from Archeops, an experiment
designed to obtain large sky coverage in a single balloon flight. A
detailed description of the instrument inflight performance will be
given in~\cite{benoit_instrument}; here we provide only essential
information. Archeops\footnote{see {\tt http://www.archeops.org}.} is
a balloon--borne experiment with a 1.5~m off--axis Gregorian telescope
and a bolometric array of 21 photometers operating at frequency bands
centered at 143~GHz (8~bolometers), 217~GHz (6), 353~GHz (6$=$3
polarized pairs) and 545~GHz (1).  The focal plane is maintained at a
temperature of~$\sim 100$mK using a $^3$He--$^4$He dilution cryostat.
Observations are carried out by turning the payload at 2~rpm producing
circular scans at a fixed elevation of $\sim 41$~deg.  Observations of
a single night cover a large fraction of the sky as the circular scans
drift across the sky due to the rotation of the Earth.

\section{Observations and processing of the data}

\begin{figure*}[!ht]
\resizebox{\hsize}{!}
{\includegraphics[clip,angle=90]{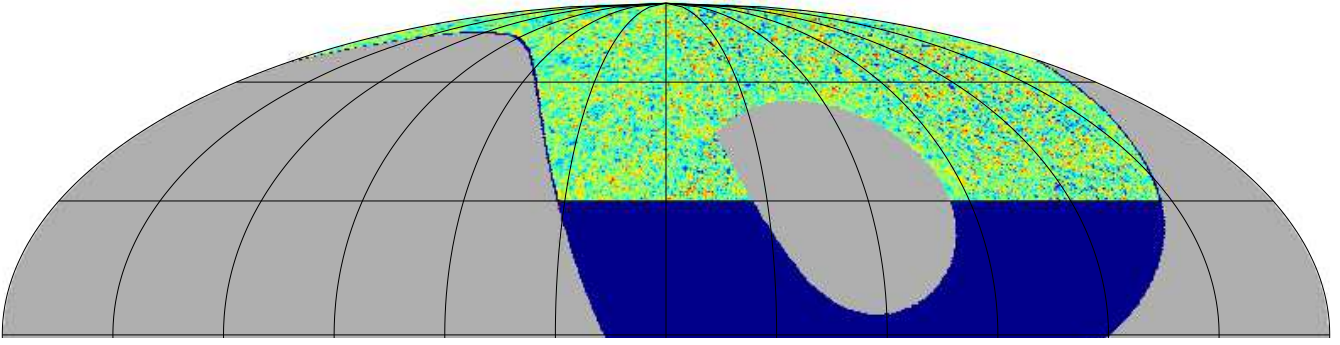}}
\caption{Archeops CMB map (Galactic coordinates, 
  centered on the Galactic anticenter, Northern hemisphere) in HEALPIX 
  pixelisation~(\cite{healpix}) with 15~arcmin. pixels and a 15
  arcmin. Gaussian smoothing. The map is a two--photometers
  coaddition.  The dark blue region is not included in the present
  analysis because of possible contamination by dust.  The colors in
  the map range from $-500$ to $500\ \mu\mathrm{K_{CMB}}$.}
\label{Fig:archcmbmap}
\end{figure*}

The experiment was launched on February 7, 2002 by the
CNES\footnote{Centre National d'{\'E}tudes Spatiales, the French
  national space agency} from the Swedish balloon base in Esrange,
near Kiruna, Sweden, $68^\circ$N, $20^\circ$E. It reached a float
altitude of $\sim 34$~km and landed 21.5 hours later in Siberia near
Noril'sk, where it was recovered by a Franco--Russian team. The
night--time scientific observations span 11.7~hours of integration
from 15.3 UT to 3.0 UT the next day.  Fig.~\ref{Fig:archcmbmap} shows
the Northern galactic part of the sky observed during the flight.

A detailed description of the data processing pipeline will be given
in~\cite{benoit_pipe}. Pointing reconstruction, good to 1~arcmin.,
is performed using data from a bore--sight mounted optical star sensor
aligned to each photometer using Jupiter observations.  The raw Time
Ordered Information (TOI), sampled at 153~Hz, are preprocessed to
account for the readout electronics and response variations. Corrupted
data (including glitches), representing less than 1.5\%, are
flagged. Low frequency drifts correlated to various templates
(altitude, attitude, temperatures, CMB dipole) are removed from the
data.  To remove residual dust and atmospheric signal, the data are
decorrelated with the high frequency photometers and a synthetic dust
timeline~(\cite{SFD}).

The CMB dipole is the prime calibrator of the instrument. The absolute
calibration error against the dipole measured by
COBE/DMR~(\cite{fixsen}) is estimated to be less than 4\% (resp.~8\%)
in temperature at 143~GHz (resp.~217~GHz).  Two other independent
calibration methods, both with intrinsic uncertainty of $\sim 10\%$,
give responsivities relative to the dipole calibration at 143
(resp.~217~GHz) of $-5$ (resp.~$+6$\%) on Jupiter and $-20$ (resp.~$-5$\%)
with COBE--FIRAS Galactic Plane emission.

The beam shapes of the photometers measured on Jupiter are moderately
elliptical, having a ratio of the major to minor axis of 1.2
(resp.~1.5) at 143~GHz (resp.~217~GHz), and have an
equivalent FWHM of 11 arcmin. (resp.~13).  The error in beam size
is less than 10\%.  The effective beam transfer function for each
photometer, determined with simulations, is taken into account in the
analysis and is in excellent agreement with analytical
estimates~(\cite{fosalba}).

\section{Analysis}

In this paper, we use data from only a single detector at each of the
CMB frequencies, 143 and 217~GHz, with a sensitivity of 90 and
150~$\mu\mathrm{K_{CMB}.s^{1/2}}$ respectively.  To avoid the
necessity of detailed modelling of Galactic foregrounds, we restrict
the sky coverage to $b > +30\rm\,deg.$, giving a total of $\sim
100,000$ 15~arcmin. pixels (HEALPIX nside~=~256) covering 12.6\% of the
sky (see Fig.~\ref{Fig:archcmbmap}).  To extract the CMB power
spectrum, we use the MASTER analysis methodology~(\cite{hivon}), which
achieves speed by employing sub--optimal (but unbiased) map--making and
spectral determinations.

First, the Fourier noise power spectrum is estimated for each
photometer.  Signal contamination is avoided by subtracting the data
projected onto a map (and then re--read with the scanning strategy)
from the initial TOI. This raw noise power spectrum is then corrected
for two important effects~(\cite{method_master}): (i)~pixelisation of
the Galactic signal that leads to an overestimate of the noise power
spectrum: sub--pixel frequencies of the signal are not subtracted from
the inital TOI leaving extra signal at high frequency; (ii)~due to the
finite number of samples per pixel, noise remains in the map and is
subtracted from the initial TOI, inducing an underestimation of the
actual noise in the final TOI (\cite{ferreira}, \cite{stompor}).
Simulations, including realistic noise, Galactic dust and CMB
anisotropies, indicate that both corrections are independent of the
shape of the true noise power spectrum, and thus permit an unbiased
estimate of the latter with an accuracy better than 1\% at all
frequencies. The corresponding uncertainty in the noise power spectrum
estimation is included in the error bars of the $C_\ell$ spectrum.

We construct maps by bandpassing the data between 0.3 and 45~Hz,
corresponding to about 30~deg. and 15~arcmin. scales, respectively.
The high--pass filter removes remaining atmospheric and galactic
contamination, the low--pass filter suppresses non--stationary high
frequency noise.  The filtering is done in such a way that ringing
effects of the signal on bright compact sources (mainly the Galactic
plane) are smaller than~$\sim 36\,\mu{\rm K}^2$ on the CMB power
spectrum in the very first $\ell$--bin, and negligible for larger
multipoles. Filtered TOI of each absolutely calibrated detector are
co--added on the sky to form detector maps. The bias of the CMB power
spectrum due to filtering is accounted for in the MASTER process
through the transfer function.  The map shown in
Fig.~\ref{Fig:archcmbmap} is obtained by combining the maps of each of
the photometers.  A $1/\sigma^2$ weighting of the data was done in
each pixel, where $\sigma^2$ is the variance of the data in that
pixel.  This map shows significant extra variance compared to the
difference map on degree angular scales which is attributed to
sky--stationary signal.

We estimate the CMB power spectrum in 16 bins ranging from $\ell=15$
to $\ell=350$. The window functions derived from the multipole binning
and renormalized to equal amplitude for clarity are shown at the
bottom of Fig.~\ref{figsyste}.  They are nearly top--hat functions due
to the large sky coverage.  The bins can therefore be approximated as
independent: off--diagonal terms in the covariance matrix are less
than $\sim 12\%$.  For the purpose of estimating the power spectrum we
made a map that combines the data of the two photometers using two
different weighting techniques. Up to $\ell=310$ the data of each
photometer has equal weight and at larger $\ell$ values the data is
noise weighted. This is valid because the multipole bins are nearly
independent. It is also advantageous because it minimizes the overall
statistical noise over the entire $\ell$ spectrum; equal weighting
gives smaller error bars at small $\ell$ and noise weighting gives
smaller error bars at large $\ell$.

\section{Results and consistency tests}

\begin{figure}[!ht]
  \resizebox{\hsize}{!}
  {\includegraphics[clip]{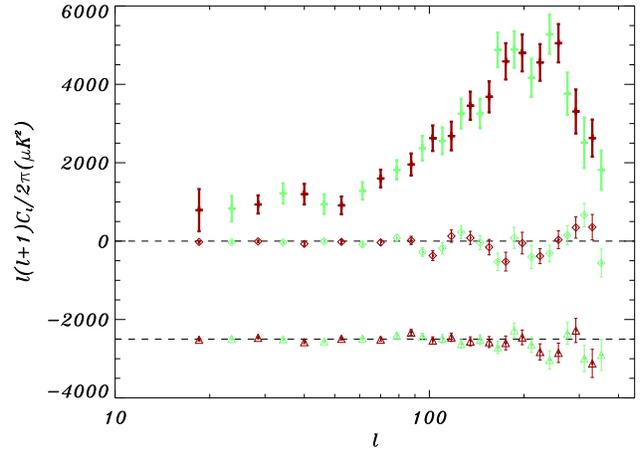}}
\caption{ The Archeops CMB power spectrum for the combination of the
two photometers. Green and red data points correspond to two
overlapping binnings and are therefore not independent. The light open
diamonds show the null test resulting from the self difference (SD) of
both photometers and the light open triangles correspond to the
difference (D) of both photometers (shifted by $-2500\ \mu\mathrm{K}^2$
for clarity) as described in sect.~\ref{sect_selfdiff} and shown in
Tab.~\ref{tab:cl}.}
\label{figarchcl}
\end{figure}

The Archeops power spectrum is presented in Fig.~\ref{figarchcl} and
in Tab.~\ref{tab:cl}. Two different binnings corresponding to
overlapping, shifted window functions (therefore not independent) were
used. Archeops provides the highest $\ell$ resolution up to $\ell=200$
($\Delta\ell$ from 7 to 25) and most precise measurement of the
angular power spectrum for $15 < \ell < 300 $ to
date. Sample--variance contributes 50\% or more of the total
statistical error up to $\ell\sim 200$.

\begin{table}[!ht]
\begin{center}
\begin{tabular}{rrr@{~$\pm$~}rr@{~$\pm$~}rr@{~$\pm$~}r}\hline\hline
$\ell_{\mathrm{min}}$ &$\ell_{\mathrm{max}}$
&\multicolumn{2}{c}{$\frac{\ell(\ell+1)C_\ell}{(2\pi)}~(\mu\mathrm{K})^2$} 
&\multicolumn{2}{c}{SD $(\mu\mathrm{K})^2$}
&\multicolumn{2}{c}{D $(\mu\mathrm{K})^2$} \\
\hline
 15 & 22 &$  789$ &$  537$ &$  -21$ &$   34$ &$  -14$ &$   34$\\
 22 & 35 &$  936$ &$  230$ &$   -6$ &$   25$ &$   34$ &$   21$\\
 35 & 45 &$ 1198$ &$  262$ &$  -69$ &$   45$ &$  -75$ &$   35$\\
 45 & 60 &$  912$ &$  224$ &$  -18$ &$   50$ &$    9$ &$   37$\\
 60 & 80 &$ 1596$ &$  224$ &$  -33$ &$   63$ &$   -8$ &$   44$\\
 80 & 95 &$ 1954$ &$  280$ &$   17$ &$  105$ &$  169$ &$   75$\\
 95 &110 &$ 2625$ &$  325$ &$ -368$ &$  128$ &$  -35$ &$   92$\\
110 &125 &$ 2681$ &$  364$ &$  127$ &$  156$ &$   46$ &$  107$\\
125 &145 &$ 3454$ &$  358$ &$   82$ &$  166$ &$  -57$ &$  114$\\
145 &165 &$ 3681$ &$  396$ &$ -154$ &$  196$ &$  -75$ &$  140$\\
165 &185 &$ 4586$ &$  462$ &$ -523$ &$  239$ &$  -97$ &$  177$\\
185 &210 &$ 4801$ &$  469$ &$  -50$ &$  276$ &$   44$ &$  187$\\
210 &240 &$ 4559$ &$  467$ &$ -382$ &$  192$ &$ -326$ &$  206$\\
240 &275 &$ 5049$ &$  488$ &$   35$ &$  226$ &$ -349$ &$  247$\\
275 &310 &$ 3307$ &$  560$ &$  346$ &$  269$ &$  220$ &$  306$\\
310 &350 &$ 2629$ &$  471$ &$  356$ &$  323$ &$ -619$ &$  358$\\
\hline
\end{tabular}
\end{center}
\caption{The Archeops CMB power spectrum for the best two photometers
(third column). Data points given in this table correspond to the red
points in Fig.~\ref{figarchcl}.  The fourth column shows the power
spectrum for the self difference (SD) of the two photometers as
described in section~\ref{sect_selfdiff}.  The fifth column shows the
power spectrum for the difference (D) between the two photometers.}
\label{tab:cl}
\end{table}

\label{sect_selfdiff}
The Archeops scanning strategy (large circles on the sky) provides a
robust test of systematic errors and data analysis procedures: by
changing the sign of the filtered TOIs every other circle, a TOI that
should not contain any signal is obtained once it is projected on the
sky. This TOI has the same noise power spectrum as the original
one. This null test is referred to as the self--difference (SD)
test. The angular power spectrum of such a dataset should be
consistent with zero at all multipoles because successive circles
largely overlap. This test has been performed with the two photometers
independently. The spectra are consistent with zero at all modes:
$\chi^2/\mathrm{ndf}$ of 21/16 (resp.~27/16) at 143~GHz
(resp.~217~GHz).  Performed on the two--photometers co--added map, the
same test gives a power spectrum consistent with zero, with a
$\chi^2/\mathrm{ndf}$ of 25/16 (see Fig.~\ref{figarchcl}). These
results show that there is no significant correlated noise among the
two photometers and that the noise model is correct. They limit the
magnitude of non--sky--stationary signals to a small fraction of the
sky--stationary signal detected in the maps.

\begin{figure}[!ht]
  \resizebox{\hsize}{!}
  {\includegraphics[clip]{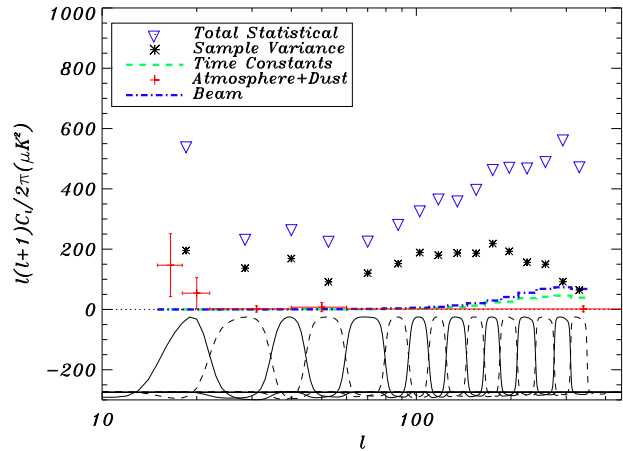}}
\caption{Contamination by systematics : the Archeops CMB power
spectrum statistical error bars (including noise and sample variance)
are shown as the blue triangles. The large error bar in the first bin
mainly comes from the high--pass filtering. A conservative
upper--limit to contamination by dust and atmospheric signal is shown
in red crosses, with a $\ell$ different binning to enhance the low
$\ell$ side. Beam and time constants uncertainties are shown in
dot--dashed blue and dashed green (see text).  The 7\% temperature
calibration uncertainty is not shown here. The window functions are
shown at the bottom of the figure.}
\label{figsyste}
\end{figure}

A series of Jack--knife tests shows agreement between the first and
second halves of the flight (the difference of the power spectra has
$\chi^2/\mathrm{ndf}=21/16$), left and right halves of the map
obtained with a cut in Galactic longitude
($\chi^2/\mathrm{ndf}=15/16$).  Individual power spectra of the two
photometers agree once absolute calibration uncertainties are taken
into account.  The power spectrum measured on the differences (D)
between the two photometers is consistent with zero with a
$\chi^2/\mathrm{ndf}$ of 22/16 (Fig.~\ref{figarchcl}) showing that the
electromagnetic spectrum of the sky--stationary signal is consistent
with that of the CMB.  The measured CMB power spectrum depends neither
on the Galactic cut (20, 30 and 40 degrees north from the Galactic
plane), nor on the resolution of the maps (27, 14 and 7' pixel size)
nor on the TOI high--pass filtering frequencies (0.3, 1 and 2~Hz).

Several systematic effects have been estimated and are summarized in
Fig.~\ref{figsyste}, along with the statistical errors (blue
triangles).  The high frequency photometer (545~GHz) is only sensitive
to dust and atmospheric emission, and thus offers a way to estimate
the effect of any residual Galactic or atmospheric emission.
Extrapolation of its power spectrum using a Rayleigh--Jeans spectrum
times a $\nu^2$ emissivity law between 545 and 217~GHz and as $\nu^0$
between 217 and 143~GHz gives an upper--limit on the possible
contamination by atmosphere (dominant) and dust. The combination of
both is assumed to be much less than 50\% of the initial contamination
after the decorrelation process.  The subsequent conservative
upper--limit for dust and atmosphere contamination is shown in
red crosses in Fig.~\ref{figsyste}.  The contamination appears
negligible in all bins but the first one ($\ell=15$ to 22).  High
frequency spectral leaks in the filters at 143 and 217~GHz were
measured to give a contribution less than half of the above
contamination.  In the region used to estimate the CMB power spectrum
there are 651~extragalactic sources in the Parkes--MIT--NRAO
catalog. These sources are mainly AGN, and their flux decreases with
frequency.  We have estimated their contribution to the power spectrum
using the WOMBAT tools~(\cite{sokasian}). At 143 (resp. 217) GHz this
is less than 2 (resp. 1) percent of the measured power spectrum at
$\ell\sim 350$.  The beam and photometer time constant uncertainties
were obtained through a simultaneous fit on Jupiter crossings.  Their
effect is shown as the dot--dashed blue and green--dashed lines in
Fig.~\ref{figsyste}.  The beam uncertainty includes the imperfect
knowledge of the beam transfer function for each photometer's
elliptical beam.  Beam and time constants uncertainties act as a
global multiplicative factor, but in the figure we show the $1\sigma$
effect on a theoretical power spectrum that has a good fit to the
data. After the coaddition of the two photometers, the absolute
calibration uncertainty (not represented in Fig.~\ref{figsyste}) is
estimated as 7\% (in CMB temperature units) with Monte--Carlo
simulations.

As a final consistency test, the Archeops $C_\ell$ are computed using
two additional independent methods.  The first is based on noise
estimation with an iterative multi--grid method,
MAPCUMBA~(\cite{mapcumba}), simple map--making and $C_\ell$ estimation
using SpICE~(\cite{spice}) which corrects for mask effects and noise
ponderation through a correlation function analysis. The second is
based on MIRAGE iterative map--making~(\cite{mirage}) followed by
multi--component spectral matching~(\cite{blindsep1,blindsep2,blindsep3}). All
methods use a different map--making and $C_\ell$ estimation.  Results
between the three methods agree within less than one $\sigma$.  This
gives confidence in both the $C_\ell$ and in the upper--limits for
possible systematic errors. Table~\ref{tab:cl} provides the angular
power spectrum which is used for cosmological parameter extraction
(\cite{benoit_params}).

\begin{figure}[!ht]
  \resizebox{\hsize}{!}
  {\includegraphics[clip]{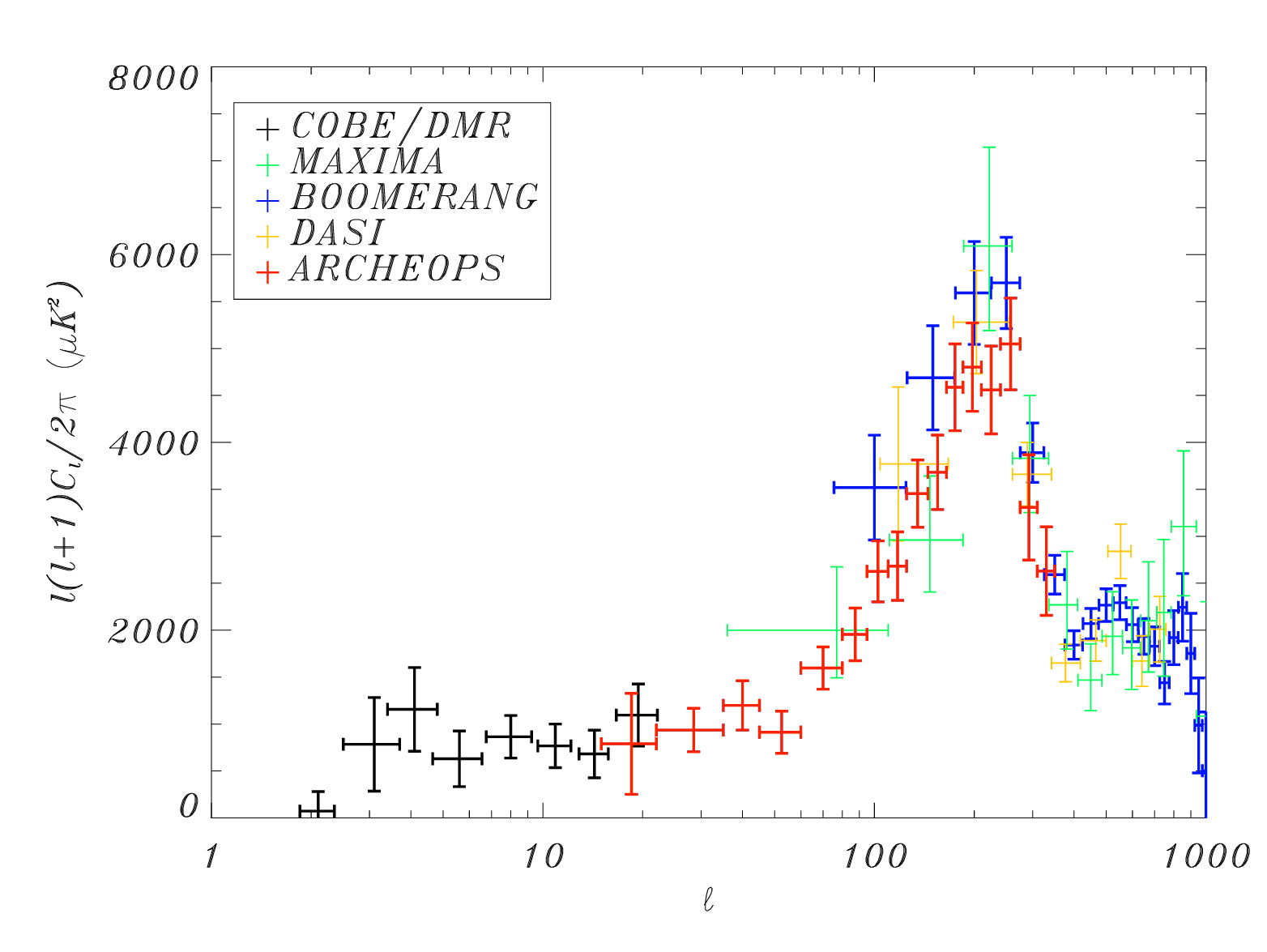}}
\caption{The Archeops power spectrum compared with 
results of COBE, Boomerang, Dasi, Maxima
 (\cite{tegmark}, \cite{boom2}, \cite{maxima2}, \cite{dasi}).
}
\label{figcompare}
\end{figure}

A comparison of the present results with other recent experiment and
COBE/DMR is shown in Fig.~\ref{figcompare}. There is good agreement
with other experiments, given calibration uncertainties, and
particularly with the power COBE/DMR measures at low $\ell$ and the
location of the first acoustic peak.  Work is in progress to improve
the intercalibration of the photometers, the accuracy and the $\ell$
range of the power spectrum: the low $\ell$ range will be improved
increasing the effective sky area for CMB (which requires an efficient
control of dust contamination), the high $\ell$ range will be improved
by including more photometer pixels in the analysis.

\section{Conclusions}

The Archeops experiment has observed a large portion of the sky.  Maps
from the two highest sensitivity detectors at 143 and 217~GHz show
consistent, sky--stationary anisotropy signal that appears
inconsistent with any known astrophysical source other than CMB
anisotropy.  The angular power spectrum of this signal at multipoles
between $\ell=15$ and $\ell=350$ shows a clear peak at $\ell\simeq
200$.  These results are consistent with predictions by
inflationary--motivated cosmologies. Archeops provides the highest
signal--to--noise ratio mapping of the first acoustic peak and its
low--$\ell$ side of any experiment to date and covers the largest
number of decades in $\ell$. It has been obtained with a limited
integration time (half a day) using a technology similar to that of
the Planck HFI experiment. An extensive set of tests limits the
contribution of systematic errors to a small fraction of the
statistical and overall calibration errors in the experiment. More
data reduction is under way to increase the accuracy and $\ell$ range
of the power spectrum.  The determination of cosmological parameters
are discussed in a companion paper~(\cite{benoit_params}).

\begin{acknowledgements}
  The authors would like to thank the following institutes for funding
  and balloon launching capabilities: CNES (French space agency), PNC
  (French Cosmology Program), ASI (Italian Space Agency), PPARC, NASA,
  the University of Minnesota, the American Astronomical Society and a
  CMBNet Research Fellowship from the European Commission.  Healpix
  package was used throughout the data analysis~(\cite{healpix}).
\end{acknowledgements}


\end{document}